\documentclass[12pt]{article}
\title{On high-redshift quasar absorption spectra and the 
Riemannian geometry of the Universe}
\author{D.~PALLE \\
Zavod za teorijsku fiziku \\
Institut Rugjer Bo\v skovi\'c \\
Po\v st. Pret. 180, HR-10002 Zagreb, CROATIA}
\begin{document}

\maketitle

\begin{abstract}
We study the observed small deviations of high-redshift quasar
absorption spectra that are interpreted as a possible 
evidence for a variable fine structure constant.
On the contrary, we claim that the effect could be
completely attributed to the small amount of cosmic
shear beyond the standard Friedmann expanding Universe.
\end{abstract}

Owing to the rapid increase of the high-resolution spectroscopy
of QSO absorption systems and the novel many-multiplet
method that allows an order of magnitude greater precision
compared with the previous alkali-doublet method \cite{webb1,webb2},
we are witnessing analyses of high-redshift quasar spectra to
the unprecedented level of precision.
The most intriguing result of these analyses is the claim of
the 4.1$\sigma$ evidence for a redshift variation of
the fine structure constant inferred from small deviations
of spectra.

In the literature one can find a lot of speculative ideas
with new particles or extra dimensions (see references
in \cite{webb2}) that produce the varying $\alpha_{e}$.
In this work we adopt a rather conservative wisdom
considering $\alpha_{e}$ constant at any redshift 
and try to explain small deviations of spectra as
a consequence of the Riemannian geometry of the Universe
beyond the Friedmann expansion.
The $\alpha_{e}(0)$ at zero momentum transfer is 
a dimensionless coupling constant
directly measurable in scattering and spectroscopic
experiments. 
This fact guided Armand Wyler to calculate $\alpha_{e}$
starting from the conformal group and symmetry \cite{wyler}
as a symmetry of Maxwell equations:

\begin{eqnarray*}
\alpha_{e}(0) = \frac{9}{8\pi^{4}}(\frac{\pi^{5}}{2^{4}5!})
^{\frac{1}{4}} .
\end{eqnarray*}

His work was strongly criticised \cite{phytoday} showing 
that one can find other similar expressions for $\alpha_{e}$. 
We can make a comment that one can actually find similar expressions
for any real number but it would be necessary to understand
a derivation of Wyler's formula with special attention to
the conformal symmetry and the dimensionality of the
physical spacetime.

For example, the conformal unification scheme for strong and 
electroweak interactions \cite{palle1} can
explain the appearance and breaking
of discrete, gauge and conformal symmetries
in a natural way.
This approach can strengthen our belief that $\alpha_{e}(0)$ is constant
in any local frame irrespective of the cosmic
distance (redshift).

The existence of cold and hot dark matter as spinning
particles \cite{palle2} and the role of
conformal symmetry at the largest distances within the
Einstein-Cartan gravity reveals the reduction of
Newton's gravitational coupling constant to the dimensionless
number of order unity \cite{palle3}:

\begin{eqnarray*}
G_{N}\rho_{\infty}H_{\infty}^{-2} = \frac{3}{4\pi} .
\end{eqnarray*}

We choose the redshift distortion as a link between two
different interpretations of measurements. 
Acknowledging the standard relation for the redshift:

\begin{eqnarray*}
\frac{\omega_{E}}{\omega_{R}} = 1 + z, 
\end{eqnarray*}
\begin{eqnarray*}
\omega_{E} (\omega_{R}) \equiv emitted (received)\ frequency, 
\end{eqnarray*}

the interpretation of Refs.\cite{webb1,webb2} could then 
be described as

\begin{eqnarray}
\frac{\tilde{\omega}_{E}}{\omega_{R}(G)} - \frac{\omega_{E}}
{\omega_{R}(G)} = \tilde{z}(G) - z(G) \equiv \delta z(\alpha_{z}),
\end{eqnarray}
\begin{eqnarray*}
G \equiv geometry\ defined\ by\ Hubble's\ expansion .
\end{eqnarray*}

Our interpretation relies on the modified Riemannian geometry of
the expanding Universe that also contains vorticity, acceleration and
shear \cite{ehlers}:

\begin{eqnarray}
\frac{\omega_{E}}{\omega_{R}(\tilde{G})} - 
\frac{\omega_{E}}{\omega_{R}(G)} = z(\tilde{G}) - z(G)
\equiv \delta z(\tilde{G}),
\end{eqnarray}
\begin{eqnarray*}
\tilde{G} \equiv H + \omega + shear + accel.
\end{eqnarray*}

Assuming small perturbations of the Friedmann metric,     
following Sachs and Wolfe \cite{sachs} one obtains

\begin{eqnarray*}
ds^{2} = R^{2}(\eta)[\eta_{\mu\nu}+h_{\mu\nu}]dx^{\mu}dx^{\nu},
\end{eqnarray*}
\begin{eqnarray*}
dt = R(\eta) d\eta,\ \eta=conformal\ time,
\end{eqnarray*}
\begin{eqnarray}
\delta z(\tilde{G}) = (z+1)\int^{\eta_{R}-\eta_{E}}_{0}
[\frac{1}{2}\frac{\partial h_{ij}}{\partial \eta}e^{i}e^{j}-
\frac{\partial h_{i0}}{\partial \eta}e^{i}]dy,
\end{eqnarray}
\begin{eqnarray*}
\mu = 0,1,2,3;\ i=1,2,3; \eta_{\mu\nu}=diag(+1,-1,-1,-1).
\end{eqnarray*}

The metric of Obukhov and Korotky \cite{obukhov} contains 
everything we need:

\begin{eqnarray*}
ds^{2} = dt^{2} - R^{2}(t)(dx^{2}+ka^{2}(x)dy^{2})
-r^{2}(t)dz^{2}-2R(t)b(x)dydt,
\end{eqnarray*}
\begin{eqnarray*}
b(x)=\sqrt{\chi}a(x),\ a(x)=Ae^{mx},\ \chi ,A,m,k=const.
\end{eqnarray*}

From Eq.(3) we conclude that only shear contributes to
the redshift distortion (it is assumed that the 
acceleration parameter $\chi$ does not depend on time):

\begin{eqnarray*}
k=1,\ a(x)=1,\ h_{zz}=-\frac{r^{2}(\eta)-R^{2}(\eta)}
{R^{2}(\eta)},\ h_{y\eta}=-\sqrt{\chi},
\end{eqnarray*}
\begin{eqnarray*}
assume:\ \frac{\dot{r}}{r}-\frac{\dot{R}}{R}=\lambda 
\frac{\dot{R}}{R},\ \dot{R}\equiv \frac{dR}{dt}, 
\lambda =const. << 1,\ r(t)\simeq R(t),
\end{eqnarray*}
\begin{eqnarray}
\delta z(\tilde{G}) = -\frac{\lambda}{3}(z+1)\ln (z+1) .
\end{eqnarray}

Acknowledging the results of Ref.\cite{webb2},
we make an order of magnitude estimate of the shear anisotropy
parameter $\lambda$:

\begin{eqnarray}
\delta z (\alpha_{z}) = \delta z (\tilde{G}),
\end{eqnarray}
\begin{eqnarray*}
\delta \omega=q_{1}x+q_{2}y \simeq (2q_{1}+4q_{2})\frac{\delta 
\alpha_{e}}{\alpha_{e}}\simeq -|{\cal O}(10^{-2})| cm^{-1},
\end{eqnarray*}
\begin{eqnarray*}
\omega \simeq {\cal O}(10^{4}) cm^{-1},\ \bar{z} = 1.5,
\end{eqnarray*}
\begin{eqnarray}
\Rightarrow \lambda = +|{\cal O}(10^{-6})|.
\end{eqnarray}

It is also observed \cite{webb2} that the effect of 
the redshift distortion grows with redshift and this 
is in accord with our Eq.(4).
One can study the evolution of shear within the fluid-flow
approach of Hawking \cite{hawking} adding the
shear viscosity term to the energy-momentum tensor
\cite{weinberg} as a consequence of the radiation-matter
coupling in primordial plasma and as a source of shear anisotropy.

Similarly, studying primordial mass density fluctuations or the
angular momentum of galaxies, one can deduce the parameters
of vorticity and acceleration \cite{palle4}.
 
Recent supernova searches and their implications on the value of
the cosmological constant give controversial results at
the redshift $z > 1$ (data are only marginally consistent
with the positive cosmological constant).
 The role of hot dark matter (light
neutrinos) is possibly underestimated in current cosmological
fits of data. Small-scale simulations within the standard
cold dark matter scenario cannot explain all observed
features.

The present complex situation concerning cosmological 
parameters could be resolved by more precise data 
(SDSS, MAP mission, etc.)
and more sophisticated cosmological models.
\newline

\hspace{50 mm}* * *
\newline
This work was supported by the Ministry of Science and Technology
of the Republic of Croatia under Contract No. 00980103.

\end{document}